\documentclass[showpacs,preprintnumbers,amsmath,amssymb,prl,twocolumn]{revtex4}
\usepackage{amssymb}
\usepackage{amsfonts}
\usepackage{mathrsfs}
\usepackage{graphicx}% Include figure files
\usepackage{dcolumn}% Align table columns on decimal point
\usepackage{bm}% bold math

\begin{document}

\title{Laser and Microwave Excitations of Rabi Oscillations of a Single Nitrogen-Vacancy Electron Spin in Diamond
}

\author{Chunyang Tang,$^{1,2}$ Xin Hu,$^{1,3}$ and Xinyu Pan$^{1,*}$}

\affiliation{%
$^1$Institute of Physics, Chinese Academy of Sciences, Beijing
100190, China \\
$^2$School of Physics, Peking University, Beijing 100871, China \\
$^3$School of Physics, Nankai University, Tianjin 300071, China
}%

%\date{\today}% It is always \today, today,
             %  but any date may be explicitly specified

\begin{abstract}

A collapse and revival shape of Rabi oscillations of a single
Nitrogen-Vacancy (NV) center electron spin has been observed in
diamond at room temperature. Because of hyperfine interaction
between the host $^{14}$N nuclear spin and NV center electron spin,
different orientation of the $^{14}$N nuclear spin leads to a
triplet splitting of the transition between the ground
\emph{m$_{s}$}=0 and excited states \emph{m$_{s}$}=1. Microwave can
excite the three transitions equally to induce three independent
nutations and the shape of Rabi oscillations is a combination of the
three nutations. This result provides an innovative view of electron
spin oscillations in diamond.

\end{abstract}

% PACS, the Physics and Astronomy
                             % Classification Scheme.
%\keywords{Suggested keywords}%Use showkeys class option if keyword
                              %display desired
\maketitle

%\section{\label{sec:level1}First-level heading:\protect\\ The line
%break was forced \lowercase{via} \textbackslash\textbacksl

{\label{sec:level1}}Manipulating single spins in solids is a
promising candidate for basic quantum computing implementation
\cite{1,2,3}. The Nitrogen-Vacancy (NV) center in diamond is unique
because its electron spin can be polarized and readout optically.
More importantly it exhibits extremely long coherence time under
optical excitation at room temperature \cite{4,5,6,7}. A quantum
logical NOT and a conditional two-qubit gate (CROT) using a nearby
$^{13}$C spin has been presented \cite{8}. Since the first
observation of coherently driven electron spin oscillations (Rabi
oscillations) of NV center was reported in 2004 \cite{9}, several
groups have repeated Rabi oscillations measurement as a basic
quantum bit manipulation \cite{10,11,12}. Collapse and revival shape
of the Rabi oscillations has been observed in two different works:
one is in type Ib diamond with rich nitrogen impurities environment
\cite{11}, the other is in type IIa diamond with a nearby $^{13}$C
nuclear spin \cite{13}. In this letter we report our experiments in
type IIa diamond (without nearby $^{13}$C and very low nitrogen
impurity content). We demonstrate a collapse and revival shape of
electron spin Rabi oscillations.

Previous theoretical analysis has studied the strong coupling to a
nearby nitrogen spin \cite{6,14} as well as nitrogen impurity spin
bath \cite{11}. We study the third regime: only taking hyperfine
coupling to the NV center's host $^{14}$N nuclear spin into
consideration. This is based on low nitrogen content in type IIa
diamond. This treatment yields a good explanation for collapse and
revival shape observed.

A NV center comprises a substitutional nitrogen atom instead of a
carbon atom and an adjacent lattice vacancy. Experiments are carried
out in a type IIa bulk diamond (high temperature and high pressure
diamond from Sumitomo Electric Industries) with extremely low
nitrogen density ($\sim$ 1 ppm ). We use a home-built laser scanning
confocal microscope system to locate the single NV centers [Fig.
1(a)]. Second order photon correlation function $g^2(\tau)$ of
center A indicates that it is a single quantum emitter [Fig. 1(b)].
All our experiments are performed at NV center A.

\begin{figure}
\includegraphics[width=0.5\textwidth]{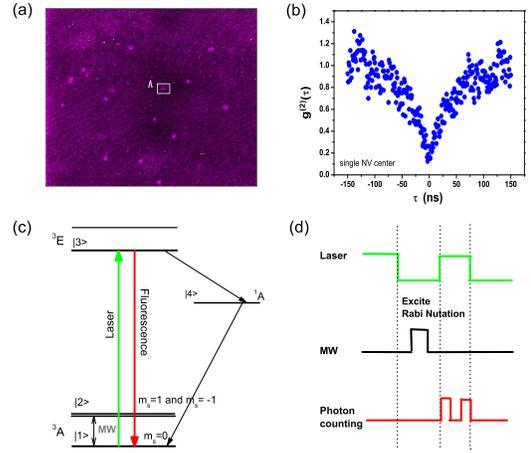}
\caption{(color online) Optical properties of NV center and Rabi
oscillations pulse sequence. (a) Two-dimension scanning confocal
image of the sample. Bright spot A corresponds to the NV center we
investigate (in all our experiments). (b) Second order photon
correlation function $g^2(\tau)$. (c) Energy level scheme of the NV
center in diamond. (d) Pulse sequence for Rabi oscillations
experiment. Optically pumped for about 3 $\mu$s, the NV center is
polarized into \emph{m$_{s}$}=0 sublevel. Microwave manipulation is
used to drive Rabi oscillations. MW duration step is 25 ns, maximum
duration is about 3.5 $\mu$s. }
\end{figure}

The electronic paramagnetic ground state of NV center is a spin
triplet state ($^3$A, \emph{S}=1 ). Fig. 1(c) shows the energy level
scheme of NV center. As a result of crystal field \cite{11}, there
is a zero field splitting \emph{D}=2.87 GHz  between sublevels of
\emph{m$_{s}$}=0 and \emph{m$_{s}$} =$\pm$1(\emph{m$_{s}$} is the
projection of spin operator along the z axis). Due to
\emph{C$_{3v}$} symmetry, \emph{m$_{s}$}=1 and \emph{m$_{s}$}=-1
levels are degenerated.

Strong optical transition between the ground states and the first
excited state (also a spin triplet state $^3$E) is dipole-allowed
and the Zero Phonon Line (ZPL) is exhibited at 637 nm (1.945 eV)
\cite{15,16}. However, fluorescence intensity of a single NV center
is strongly dependent on its spin state. Average photon emission
rate is quite smaller for transitions involving the levels
\emph{m$_{s}$} =$\pm$1 than for the level \emph{m$_{s}$}=0, which
allows polarization to \emph{m$_{s}$}=0 sublevel via optical pumping
\cite{17,18,19} and readout of the spin state by spin-selective
fluorescence \cite{20}. Spin-dependent intersystem crossing to a
metastable level (singlet, $^1$A ) is assumed to be responsible for
these differences \cite{4,21}.

In all our experiments, an external static magnetic field of $\sim$
40 Gauss is applied to split the \emph{m$_{s}$} =$\pm$1 sublevels by
$\sim$ 60 MHz [detailed electron spin resonance spectrum is shown in
Fig. 3]. We set our microwave frequency resonant to \emph{m$_{s}$}
=1 level. Fig. 1(d) is the pulse sequence for Rabi oscillations
measurement. First, the laser is turned on to pump NV center
electron spin into \emph{m$_{s}$}=0 state, then the microwave (MW)
pulse of different time duration is used to manipulate the electron
spin. The final spin state is read out by a second laser pulse.
Every cycle of the pulse sequence is about 1 ms, we repeat the
measurement cycle for 10$^5$ times to obtain a smooth oscillation
signal which takes about 1 min.

\begin{figure}
\includegraphics[width=0.5\textwidth]{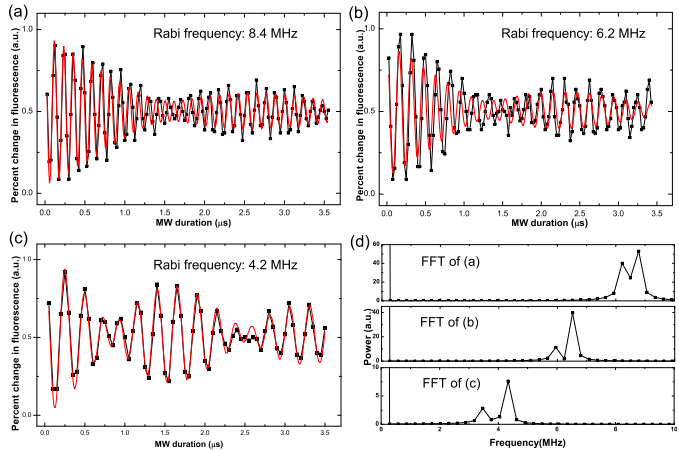}
\caption{(color online) Collapse and revival shape of Rabi
oscillations. Experimental data and calculation considering coupling
to host $^{14}$N nuclear spin for Rabi frequency of (a) 8.4 MHz, (b)
6.2 MHz and (c) 4.2 MHz. Red lines are fit to an exponentially
decayed sum of three cosines with different frequencies (see Eq. 3
in text). All curves are obtained at MW detuning $\delta_f$ = 0.
Laser power is $\sim$ 300 $\mu$W for all curves. (d) is Fast Fourier
Transform (FFT) of (a), (b) and (c). }
\end{figure}

We first show a collapse and revival shape of Rabi oscillations
[Fig. 2, black points and lines] for three Rabi frequencies 8.4 MHz,
6.2 MHz and 4.2 MHz. In all cases the coherent oscillations have an
envelope: the signal initially collapses and then revives with
decreased amplitude. Weak MW induced oscillation collapses at t $<$1
$\mu$s while strong MW leads the to a later collapse time t $\sim$
1.5 $\mu$s. Fig. 2(d) shows Fast Fourier Transform (FFT) of the
three oscillations mentioned above and indicates that all of the
three oscillations contain two different frequency components.

R. Hanson et al. \cite{11} observed a collapse and revival shape of
oscillations in type Ib diamond with rich nitrogen impurities and
regarded it as a result of effective spin bath. However, we use type
IIa diamond which contains much lower nitrogen impurities than type
Ib and it offers much cleaner spin environment. M.D. Lukin has
reported a similar shape of electron spin oscillations in NV center
in type IIa diamond and attributed it to the influence of a nearby
$^{13}$C nuclear spin \cite{13}. Our Electron Spin Resonance (ESR)
spectrum shows a clean environment around the NV center without
nearby nuclear spins [Fig. 3(a) and (b)]. A $\sim$ 60 MHz splitting
is induced by Zeeman effect. We do not observe splitting caused by
$^{13}$C. The ESR spectrum shows a 2.2 MHz triplet splitting, which
is due to the hyperfine interaction of electron spin with host
$^{14}$N nuclear spin \cite{22} [Fig. 3(b)].

\begin{figure}
\includegraphics[width=0.5\textwidth]{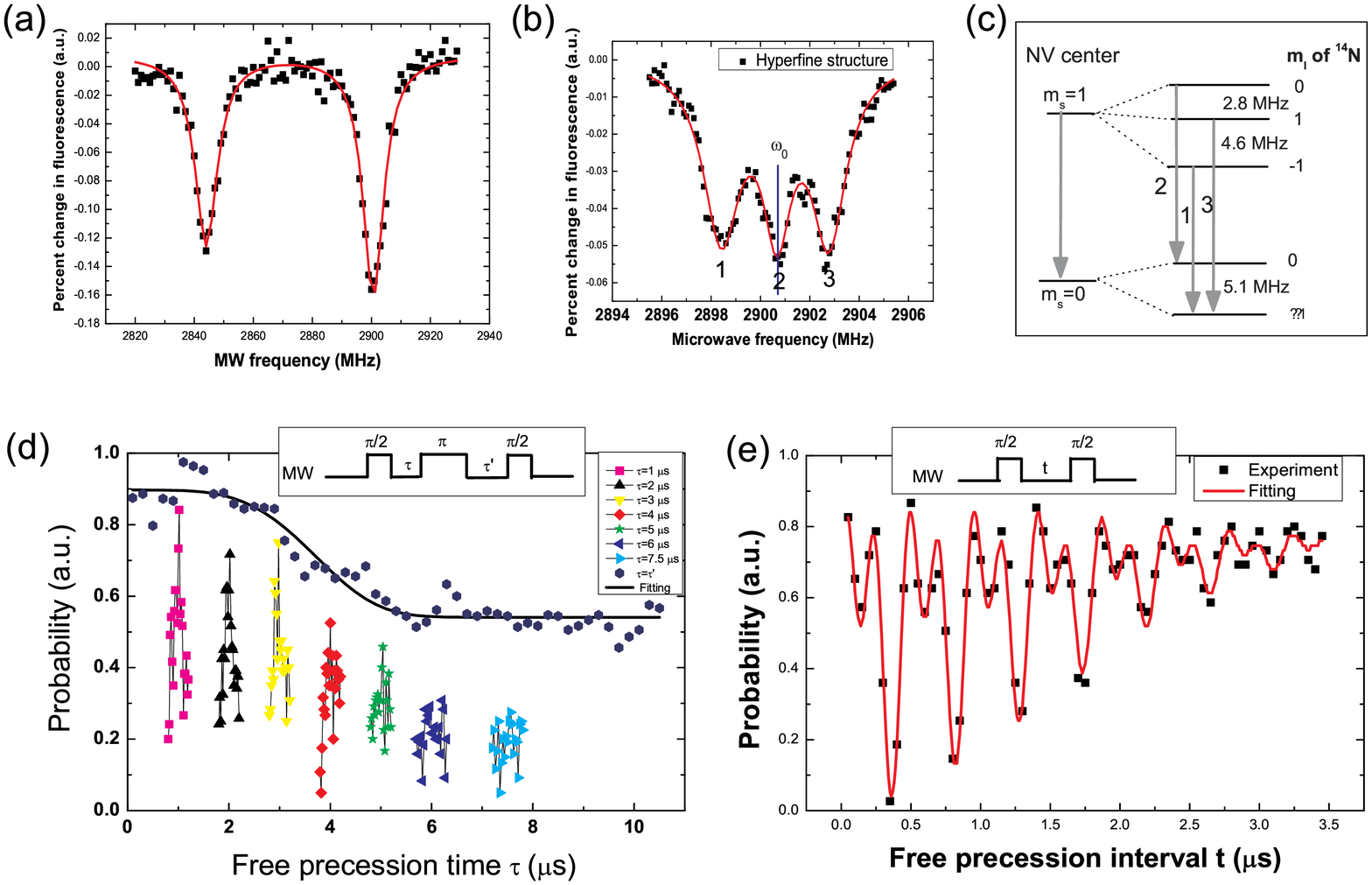}
\caption{(color online) ESR spectrum, hyperfine energy level, Ramsey
fringes and spin echo. (a) ESR spectrum of the NV center. Two main
peaks correspond to \emph{m$_{s}$}=1 and \emph{m$_{s}$}=-1. (b) ESR
under weak MW and laser power, triplet splitting of 2.2 MHz is
shown, which is an indication of the host $^{14}$N. Red curves
represent the fit in the form of sum of three Lorentz functions. (c)
Theoretical calculated hyperfine energy level scheme. The three
allowed hyperfine transitions are numbered and correspond to three
ESR signal troughs in (b). (d) Spin echo signal at low magnetic
field $\sim$ 40 Gauss. (e) Ramsey fringes signal. MW pulse sequences
for (d) and (e) are depicted as inset into each figure and resonant
Rabi frequency for spin echo and Ramsey is about 4 MHz.}
\end{figure}

To gain insight into the dynamics of NV center electron spin in
clean environment, we consider the hyperfine interaction with host
$^{14}$N nuclear spin. The external magnetic field shifted the two
\emph{m$_{s}$} =$\pm$1 transitions by $\sim$ 60 MHz, we disregard
\emph{m$_{s}$}=-1 level and treat NV center as a two-level system.
We denote \emph{m$_{s}$}=0 as level 1 and \emph{m$_{s}$}=1 as level
2. The Hamiltonian regarding this hyperfine interaction can be
written as: \cite{11}

\begin{eqnarray}
H=A_0 S^z_0 I^z_0 + A_1 (S^x_0 I^x_0 + S^y_0 I^y_0) - P_0 (I^z_0)^2
\end{eqnarray}

,where A$_0$= 2.3 MHz and A$_1$= 2.1 MHz are hyperfine coupling
tensor parameters, and P$_0$= -5.1 MHz is quadrupole splitting for
nucleus having spin I $\geq$ 1. The hyperfine energy level scheme
\cite{22,23} is shown in Fig. 3(c): there are three transitions
corresponding to three possible z projections of the $^{14}$N
nuclear spin and these transitions are separated by $\alpha_N
\approx $2.2 MHz in energy. We denote middle of the three
transitions as $\omega_0 = 2\pi f^{NV}_0$ and denote MW frequency in
detuning formula $\delta_f =f_{MW} - f^{NV}_0$.

This 2.2 MHz splitting is essentially quasi-static because of slow
nuclear spin relaxation ($\sim$ 1 s) compared with fast measurement
in a single pulse sequence. In other words, orientation of the host
nitrogen nuclear spin does not change during a single measurement
cycle. This hyperfine interaction therefore only gives an additional
detuning to the resonant MW frequency between \emph{m$_{s}$}=0 and
\emph{m$_{s}$}=1. We can still treat NV center as a two-level system
by non-resonant excitation (adiabatic approximation).

As we repeat pulse sequence for 10$^5$ cycles ($\sim$ 1 minute), the
free evolution of $^{14}$N gives rise to three possible z
projections of its nuclear spin (m$_I$ =+1,0,-1). Physically this
nuclear spin flips in space, but a static N nuclear spin model is
applicable. The three orientations of $^{14}$N nuclear spin have the
same probability 1/3. MW pulse can excite all the three transitions
and they are equally mixed. We average over all three possible Rabi
oscillations (ergodic assumption). The Rabi nutation frequencies
will arise when the $^{14}$N spin is in its three possible
projections along the NV axis. As nutations with different
frequencies superpose, a beat is present.

Rabi signal considering dephasing process (provided rotating wave
approximation is applicable) is an exponent-decayed population
change between level 1 and 2 \cite{24}:

\begin{eqnarray}
|a_1 (t) |^2 = e^{-t/t_0} \frac{f^2_0}{f^2_e} \cos 2 \pi f_e t
\end{eqnarray}

, where $|a_1 (t) |^2$ corresponds to population of level 1. t$_0$
represents decay constant. $f_0=\gamma H_R /2\pi$  is the resonant
Rabi frequency with $\gamma$ gyromagnetic ratio of electron and
$H_R$ provided by MW field. $f_e=(f^2_0+\delta ^2_f)^{1/2}$  is the
effective Rabi frequency for a microwave detuning $\delta _f$. There
exist three transitions and three detunings, respectively. Average
of the three nutations is:

\begin{eqnarray}
|a_1 (t) |^2 = \frac{1}{3} e^{-t/t_0} \sum_{f_e} \frac{f^2_0}{f^2_e}
\cos 2 \pi f_e t
\end{eqnarray}

In order to compare this model with experiment, we perform Fast
Fourier Transform (FFT) to obtain frequency information [Fig. 2(d)].
We fit the theoretical model (Eq. 3) to experiment curves [Fig. 2,
red line].

For short measurement time and large Rabi resonant frequency, it is
possible to observe oscillations with lasting decay trend because
the first part of a wave packet is just collapse of the nutation. F.
Jelezko et al. \cite{9} has reported oscillations with lasting decay
tendency. Their Rabi frequency is very high (16 MHz and 39 MHz) and
their measurement lasts about 2 $\mu$s. In contrast with their
experiments, we do not use very strong MW power to excite nutations
with Rabi frequency larger than 10 MHz and set our maximum MW
duration relatively long ($\sim$ 3.5 $\mu$s). Collapse and revival
shape should be a general form of Rabi nutations since MW can excite
all three transitions within the hyperfine structure as long as the
resonant Rabi frequency is bigger than the hyperfine splitting.

We can estimate the characteristic free precession time of $^{14}$N
. The free precession characteristic time of   should be much longer
than $\sim$ 1 ms (the time for a single pulse sequence) and should
be much shorter than $\sim$ 1 min (the time to repeat 10$^5$
cycles). This is a rough estimation and more accurate estimation may
be obtained by reducing the cycle number to determine when ergodic
assumption will not be applicable.

We also performed spin echo [Fig. 3(d)] and Ramsey fringes [Fig.
3(e)] experiment to determine the spin coherence property of the NV
center. The Ramsey fringes are obtained using a microwave pulse
sequence $\pi /2-t-\pi /2$  for a microwave detuning $\delta_f$=
-3.3 MHz. Our theoretical simulation indicates that the spin
dephasing time $T^*_2$ is about 2.3 $\mu$s. To eliminate frequency
shifts caused by slowly changing nuclear spin environment, we
utilize $\pi /2-\tau -\pi -\tau^{'} - \pi/2$ known as spin echo
technique. Fit to the envelop of the spin echo signal (peaks of the
curve $\tau = \tau^{'}$) gives the coherence time $\tau_c \approx 4
\mu s$. This coherence time for 40 Gauss external static magnetic
field is comparable with previous reported values \cite{10}.

We demonstrate the response of Rabi oscillations to different MW
frequencies [see Fig. 4]. We set the MW detunings to $\delta_f$=0,
1.1, 2.2, 3.3 MHz. Hyperfine structure plays an important role here
as in the case $\delta_f$= 0[see Fig. 2]. Frequency information from
FFT of all curves (figures not shown) is in agreement with the
effective Rabi frequency $f_e = (f^2_0 + \delta^2_f)^{1/2}$ . The
resonant Rabi frequency $f_0$ of different curves are slightly
different (error less than 5\%) and may be attributed to the MW
power fluctuation. A fit based on equal-weight combination
theoretical model is shown in Fig. 4 (red lines). These results
demonstrate that the MW frequency detuning can have an influence on
determining Rabi period: complicated curve shape may occur under
certain detunings. The resonant Rabi frequency and $\pi$ pulse can
be determined correctly only if the hyperfine interaction induced by
the host $^{14}$N nuclear spin is considered.

\begin{figure}
\includegraphics[width=0.5\textwidth]{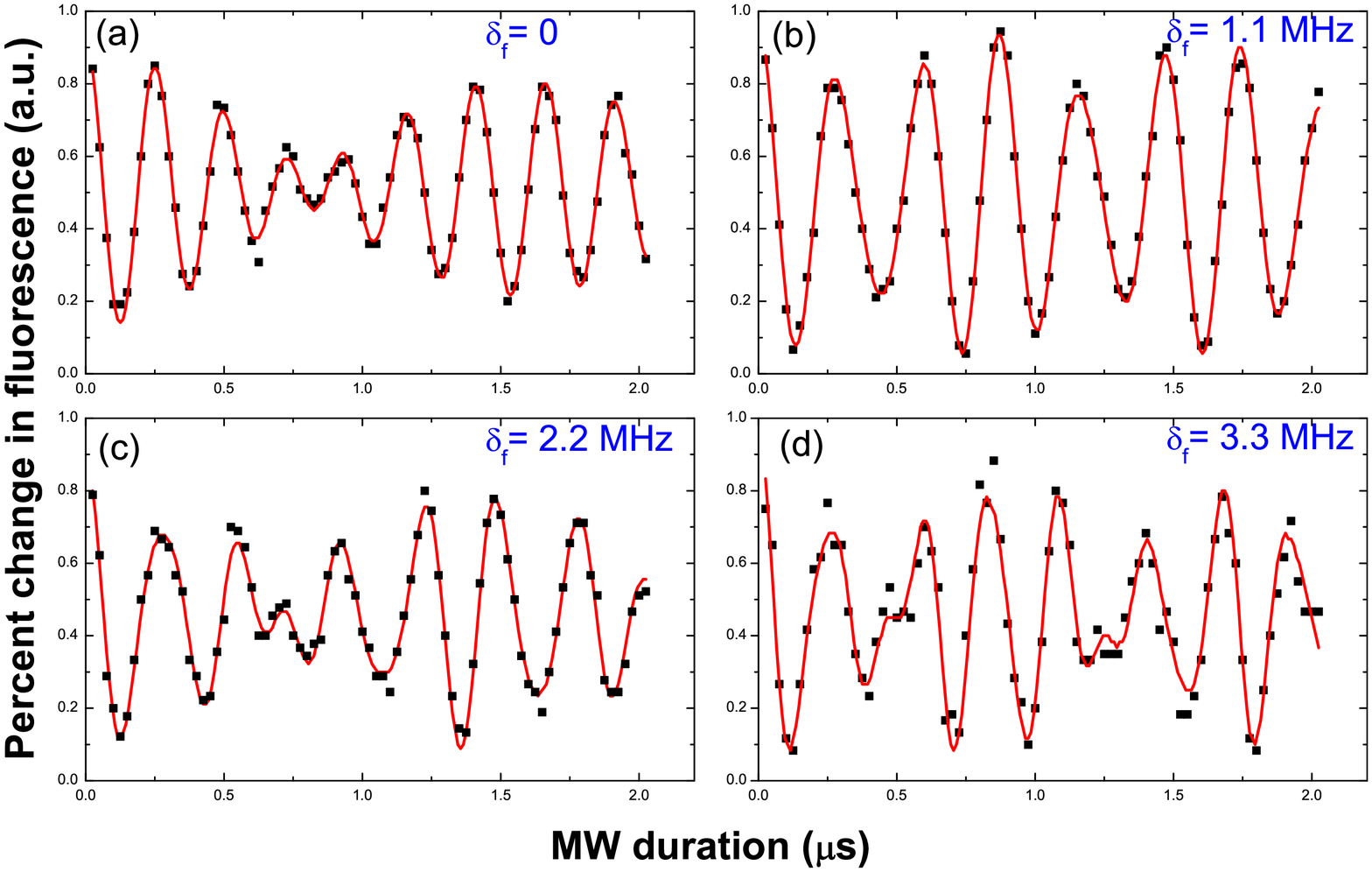}
\caption{(color online) Influence of microwave detuning on Rabi
oscillations. Rabi oscillations for detuning $\delta_f$=0, 1.1, 2.2,
3.3 MHz. All experiments are carried out under the same microwave
power. Fit to exponent decayed sum of three nutations is shown by
red line. }
\end{figure}

In summary, we have demonstrated the influence of nuclear spin
induced hyperfine interaction on Rabi oscillations. Different
orientation of the $^{14}$N nuclear spin leads to a triplet
splitting of the transition between the ground \emph{m$_{s}$}=0 and
excited \emph{m$_{s}$}=1 states. Microwave can excite the three
transitions equally to induce three independent nutations and Rabi
signal is a combination of the three nutations. The hyperfine
structure due to $^{14}$N exists whatever the NV center's
environment is.

The authors thank Lili Yang, Xinyu Luo and Wang Yao for helpful
discussions. This work was supported by National Basic Research
Program of China (973 Program project No. 2009CB929103), the NSFC
Grant 10974251.

$^*$Email: xypan@aphy.iphy.ac.cn

\newpage
%\bibliography{apssamp}
\end{document}